\documentclass[twocolumn]{aastex701}

\graphicspath{{figures/}}

\begin{document}

\title{Effects of Geomagnetic Cutoff Rigidity Variations during Forbush Decreases}

\author{Pengwei Zhao}
\affiliation{%
School of Physics and Astronomy, Sun Yat-sen University, Zhuhai 519082, China
}%
\makeatletter
\expandafter\gdef\csname email@\romannumeral\the\allauthorcnt\endcsname{}
\makeatother


\author{Jie Feng}
\affiliation{%
School of Science, Shenzhen Campus of Sun Yat-sen University, Shenzhen 518107, China
}%
\email{fengj77@mail.sysu.edu.cn}

\submitjournal{ApJL}

\begin{abstract}
Forbush decreases (FDs) are short-term reductions in galactic cosmic ray flux caused by interplanetary disturbances. During some interplanetary coronal mass ejection (ICME) events, neutron monitor (NM) data also contain variations produced by geomagnetic storms. Earlier studies emphasized apparent effects near 10~GV, but storm-time changes in geomagnetic cutoff rigidity can either increase or decrease the ground-level count rate. Using a recently published hourly proton flux reconstructed from NM data for May 2011 through October 2019, the interval covered by the published AMS daily proton fluxes, we show that these localized anomalies can extend to lower rigidities and reach 1~GV in some events. Such effects can bias the rigidity dependence inferred from NM-based hourly proton spectra during disturbed intervals. Because AMS measures proton rigidity directly in space, its daily proton spectrum is not affected by cutoff variations at ground stations and provides a stable reference. We therefore use AMS to constrain corrections for selected events. The correction removes localized anomalies while preserving the broader FD evolution, and for a representative ICME event it brings the corrected daily averages closer to the AMS measurements. Our results show that short-timescale cosmic ray variability during FDs reflects both heliospheric modulation and storm-time changes in geomagnetic shielding.
\end{abstract}

\section{Introduction}
\label{sec:introduction}

Galactic cosmic ray protons in the heliosphere vary on timescales from the solar cycle to disturbances lasting only a few hours or days. Forbush decreases (FDs) are variations on short timescales associated with interplanetary coronal mass ejections (ICMEs) and corotating interaction regions (CIRs; \citealt{1971SSRv...12..658L,2000SSRv...93...55C,2018LRSP...15....1R}). In many ICME events, the proton flux drops rapidly and then recovers more slowly. Some events also show a two-step profile when the shock and the ejecta contribute separately to the modulation. Resolving these structures requires proton flux measurements with hourly time resolution.

Ground-based neutron monitors provide continuous observations with high time resolution over many decades, but their count rates are integral responses to primary cosmic rays above station-dependent geomagnetic cutoff rigidities and therefore cannot directly provide proton fluxes in separate rigidity intervals (\citealt{2021JGRA..12628941V}). Space-based instruments can measure proton fluxes directly in rigidity bins. AMS has published daily proton fluxes from 1 to 100~GV (\citealt{2021PhRvL.127A1102A}), and these data have been used to study the rigidity dependence of FD amplitudes (\citealt{2023ApJ...950...23W}). Daily averages, however, are too coarse to resolve much of the internal evolution of FD events. Finer time resolution has been reported for other particle species, for example by DAMPE for electrons and positrons (\citealt{2021ApJ...920L..43A}), but comparable hourly proton fluxes over the rigidity range considered here are still unavailable.

During strong ICME events, neutron monitor observations can also include a geomagnetic contribution. \citet{2023JASTP.25206146G} studied several extreme ICME-driven geomagnetic storms and found count-rate enhancements during the storm main and recovery phases that were consistent with a reduction in geomagnetic cutoff rigidity. \citet{2005JGRA..110.9S20B} noted that cutoff rigidity can also increase at the start of a magnetic storm, which would temporarily reduce the cosmic ray intensity at ground level. Earlier work likewise showed that geomagnetic disturbances can affect the rigidity dependence inferred from neutron monitor data during FDs (\citealt{2013SoPh..286..561A}) and that large cutoff-rigidity variations do occur during intense storms (\citealt{2013AdSpR..51.1230T}). Part of the observed signal during some FDs may therefore reflect changing geomagnetic shielding rather than heliospheric modulation alone.

This motivates our analysis strategy. NM count rates are measured at fixed locations under geomagnetic shielding that changes with time, so storm-time cutoff variations can be folded into FD signatures inferred from ground data. AMS, by contrast, measures particle rigidity directly in space for individual events with controlled exposure. Its daily proton flux is therefore unaffected by local cutoff variations at ground stations. We use AMS as an external spectral reference to evaluate and correct NM-based hourly reconstructions during FDs.

In this paper, we focus on localized rigidity-dependent deviations that appear in NM-based hourly proton reconstructions during selected ICME events between May 2011 and October 2019, when AMS daily proton fluxes are available as an external reference. These deviations are confined in both time and rigidity, and they are difficult to explain as heliospheric modulation alone. If left uncorrected, they can distort the inferred rigidity dependence of the FD. Our goal is to identify the affected intervals and correct them with a physically motivated external spectral reference while preserving the hourly FD evolution in the unaffected bins.

Our analysis leads to two main conclusions. First, the spectral distortions previously discussed as cutoff-related effects near 10~GV also appear at lower rigidity, reaching 1~GV in some hourly NM-based reconstructions. Second, because these distortions bias the spectral behavior during events, the published hourly NM-based proton flux requires corrections in the affected rigidity bins and time intervals.

The remainder of this paper is organized as follows. Section~\ref{sec:data_products} describes the data sets and defines the target anomaly. Section~\ref{sec:ams_constraint} presents the AMS-based spectral constraint, Section~\ref{sec:correction_method} introduces the event-targeted correction method, and Section~\ref{sec:results} presents the correction results. Section~\ref{sec:discussion} summarizes the main conclusions and discusses the physical implications.

\section{Data}
\label{sec:data_products}

This analysis uses two data products: the hourly proton flux reconstructed from neutron monitor (NM) observations in our previous work (\citealt{2026ApJS..282....5Z}) and the daily proton flux measured by AMS (\citealt{2021PhRvL.127A1102A}). In that earlier study, count rates from multiple NM stations were mapped to proton fluxes in discrete rigidity intervals with a deep learning model trained on AMS daily fluxes. The same mapping was then applied to hourly NM inputs after removal of the ground diurnal component. The resulting hourly product preserves the rigidity binning of the daily reconstruction and serves here as the uncorrected series. In the present work, we restrict the analysis to May 2011 through October 2019, the interval covered by the published AMS daily proton fluxes, so that each event can be evaluated against an AMS reference spectrum.

We use the AMS daily fluxes as a spectral constraint rather than a direct replacement for the hourly reconstruction. From the AMS daily fluxes in 30 rigidity intervals between 1 and 100~GV, we derive local spectral parameters that describe the daily proton spectrum during ICME events.

AMS is adopted as the spectral reference because it measures proton flux directly in rigidity bins with a magnetic spectrometer in space, using event-by-event rigidity reconstruction and well-defined exposure selection. Unlike NM count rates, AMS measurements are effectively decoupled from the terrestrial geomagnetic cutoff at fixed ground stations. Changes in geomagnetic shielding during storms, whether they lower or raise the effective cutoff rigidity, therefore do not introduce a comparable bias into the AMS fluxes. AMS daily fluxes provide an appropriate external reference for identifying and correcting localized anomalies in the NM-based hourly reconstruction.

\begin{figure}[ht!]
\centering
\includegraphics[width=0.5\textwidth]{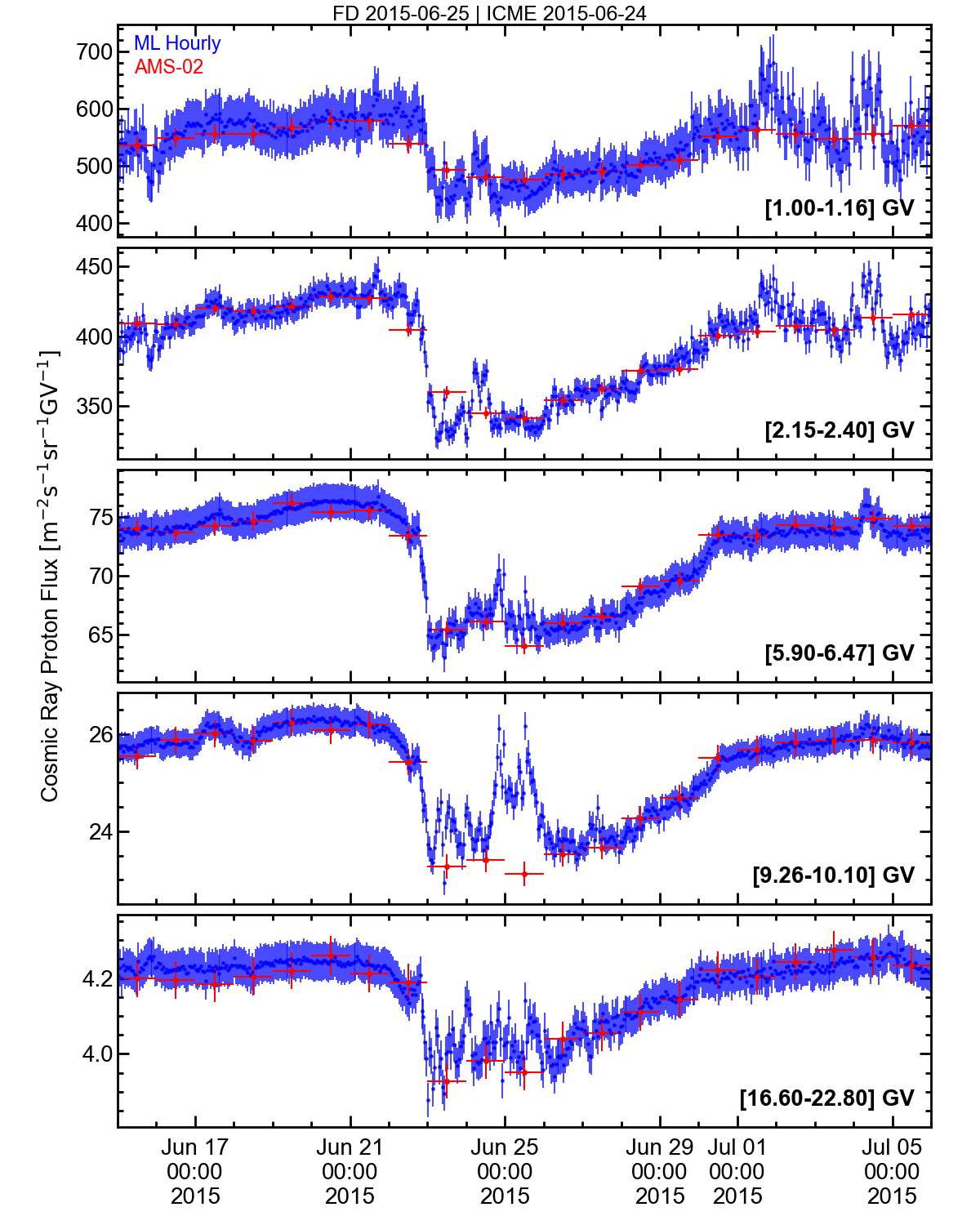}
\caption{Comparison between the uncorrected hourly reconstruction and the AMS daily flux for the 2015 June 25 ICME event. Localized deviations appear during the disturbed interval, including a representative excess above the AMS level in the [9.26, 10.1]~GV interval and a representative deficit below the AMS level in the [2.15, 2.4]~GV interval.}
\label{fig:ml_ams_comparison_FD20150625_ICME}
\end{figure}

We focus on localized deviations during selected ICME events. A representative case is shown for the 2015 June 25 event in Figure~\ref{fig:ml_ams_comparison_FD20150625_ICME}. In the [9.26, 10.1]~GV interval, the uncorrected hourly reconstruction rises above the trend defined by neighboring rigidity bins and its daily average exceeds the AMS daily flux. The same event also contains a localized deficit in the [2.15, 2.4]~GV interval, as discussed in Section~\ref{sec:results}. Consistent with earlier NM studies that linked FD-time anomalies to storm-time cutoff-rigidity variations (\citealt{2013SoPh..286..561A,2023JASTP.25206146G,2005JGRA..110.9S20B}), we interpret these deviations mainly as geomagnetic-cutoff effects rather than failures of the machine-learning mapping. The AMS daily proton data, however, show that FD days and quiet days follow the same overall parameter--flux trends in the appendix analysis. This suggests that the main issue in the hourly reconstruction is a localized distortion superposed on an otherwise similar day-scale spectral trend. We identify the affected bins through daily comparisons with AMS within each event window.

\section{Spectral Constraint}
\label{sec:ams_constraint}

We use the AMS daily proton spectrum as an external constraint on the event-scale spectral shape during ICME intervals. Following the local-slope analysis described in Appendix~\ref{app:ams_constraint_details}, we parameterize the daily spectral shape as
\begin{equation}
\gamma(R)=\gamma_{\infty}+\frac{A}{R^{k}},
\label{eq:gamma_fit}
\end{equation}
where $\gamma_{\infty}$ is the high-rigidity limit, $A$ sets the low-rigidity deviation, and $k$ controls how fast that deviation decreases with rigidity.

For the main analysis, the key result is that the AMS-inferred spectral shape changes only modestly on short timescales during the selected ICME intervals, compared with the localized anomalies seen in the NM-based hourly reconstruction. We therefore treat the AMS daily spectral fit as a stable short-timescale constraint for identifying and correcting affected NM bins.

For each ICME event, we select bins within the detection window when the absolute daily-average deviation from AMS exceeds the AMS uncertainty at the $3\sigma$ level. Each continuous candidate interval is then extended by one day on both sides. Outside the selected bins and dates, the original hourly reconstruction is retained.

\section{Correction Method}
\label{sec:correction_method}

We use the AMS daily spectral fit to describe the rigidity dependence during each corrected interval. By integrating Eq.~(\ref{eq:gamma_fit}), we obtain

\begin{equation}
\Phi(R,t)=C_t\,R^{\gamma_{\infty}(t)}
\exp\left[-\frac{A(t)}{k(t)}R^{-k(t)}\right],
\label{eq:flux_shape}
\end{equation}

where $C_t$ is the hourly normalization coefficient. We assign the daily parameters $(\gamma_{\infty}, A, k)$ to noon on each day and linearly interpolate them to hourly resolution. This interpolation provides a continuous spectral constraint across the event without directly inserting AMS daily fluxes into the hourly series.

We apply the correction separately to each ICME event and only to the bins and dates selected in Section~\ref{sec:ams_constraint}. At each corrected hour, we determine $C_t$ from the reconstructed fluxes in the remaining rigidity bins at that same time. We then evaluate the corrected flux in the target bin with the same spectral form,

\begin{equation}
\Phi^{\rm recon}_t=C_t\,R_{\rm tar}^{\gamma_{\infty}(t)}
\exp\left[-\frac{A(t)}{k(t)}R_{\rm tar}^{-k(t)}\right].
\label{eq:phirecon}
\end{equation}

where $R_{\rm tar}$ denotes the representative rigidity of the selected bin. When the reconstruction is available, we replace the original hourly value in that bin with $\Phi^{\rm recon}_t$. All other bins and times remain unchanged. In this procedure, AMS enters only through the interpolated spectral parameters and the selection of bins. The hour-to-hour normalization is still set by the NM-based reconstruction in the remaining bins. The corrected product should therefore be interpreted as an hourly reconstruction constrained by the daily AMS spectral fit, rather than as an hourly AMS measurement.

The uncertainty of the corrected flux is obtained from the fit uncertainty of the spectral slope. In practice, we estimate the uncertainty of the fitted $\gamma(R)$ relation and propagate it through Eq.~(\ref{eq:phirecon}) to obtain the error assigned to $\Phi^{\rm recon}_t$.

\section{Results}
\label{sec:results}

\begin{figure*}[t!]
\centering
\includegraphics[width=\linewidth]{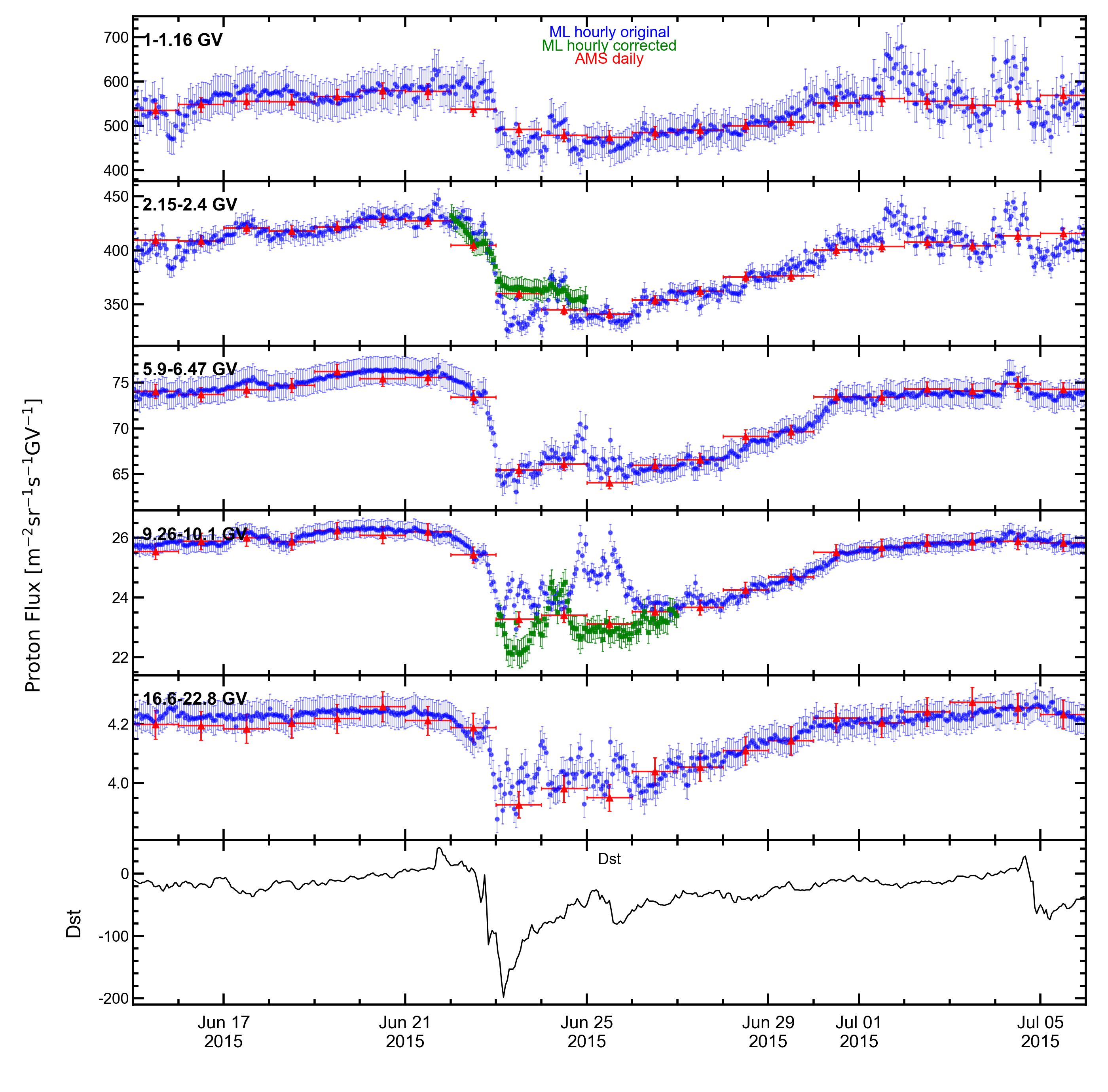}
\caption{Correction result for the ICME event of 2015 June 25. The panels show five representative rigidity bins together with the Dst index. Blue, red, and green points denote the original hourly reconstruction, the AMS daily flux, and the corrected hourly values, respectively. The [2.15, 2.4]~GV and [9.26, 10.1]~GV intervals are shown as representative corrected bins. In the lower-rigidity interval, the original reconstruction lies below the AMS daily flux; in the higher-rigidity interval, it lies above the AMS level. In both cases, the corrected series moves toward the AMS daily flux.}
\label{fig:FD20150625_time_profile}
\end{figure*}

Figure~\ref{fig:FD20150625_time_profile} shows the original and corrected hourly reconstructions in five representative rigidity bins for the 2015 June 25 ICME event, together with the AMS daily flux and the disturbance storm time (Dst) index (\citealt{2020hdrl.data....1P}). Among the bins shown, the [2.15, 2.4]~GV and [9.26, 10.1]~GV intervals provide representative examples of the corrected deviations. In the lower-rigidity bin, the original hourly reconstruction falls below the AMS daily flux during the disturbed interval, whereas in the higher-rigidity bin it exceeds the AMS level. After correction, both bins move closer to the AMS daily flux. The other three bins are shown for comparison.

The Dst index provides the geomagnetic context of the event. The largest deviations of the original hourly NM-based reconstruction from the AMS daily reference occur during the interval of most negative Dst, when storm-time changes in the effective geomagnetic cutoff rigidity are expected to be strongest. We use Dst here only as an indicator of geomagnetic disturbance, not as a direct measurement of cutoff rigidity. The figure therefore does not imply a one-to-one relation between Dst and the flux deviation in each rigidity bin. It does show, however, that the strongest departures from the AMS-constrained behavior coincide with the period when geomagnetic shielding is likely to change most rapidly.

The two corrected bins illustrate the effect clearly. In the [2.15, 2.4]~GV interval, the correction raises the hourly reconstruction toward the AMS level. In the [9.26, 10.1]~GV interval, it lowers the reconstruction toward AMS. In both bins, the corrected daily averages move closer to unity when normalized to the AMS daily flux.

For this event, the correction is limited to a small set of selected day--bin combinations and leaves the other bins unchanged. The result is a local adjustment of the spectrum during the disturbed interval rather than a global modification of the hourly reconstruction.

\section{Conclusions}
\label{sec:discussion}

Within the interval from May 2011 through October 2019, we identify localized deviations in the NM-based hourly proton reconstruction during selected ICME events and correct them with a daily spectral constraint from AMS. For each event, we compare daily averages with AMS, select bins for which the absolute deviation exceeds the AMS uncertainty at the $3\sigma$ level, extend each continuous sequence by one day on both sides, and reconstruct the selected hourly values with interpolated AMS spectral parameters and the NM-based normalization from the remaining bins.

The correction remains local in both rigidity and time. In the representative event of 2015 June 25, the [2.15, 2.4]~GV and [9.26, 10.1]~GV intervals provide clear examples of the corrected deviations during the disturbed interval. The anomalies are confined to limited parts of the spectrum rather than spread across the full reconstruction.

These results show that the spectral distortions discussed in NM-based FD studies as cutoff-related effects near 10~GV can also appear at lower rigidity, reaching 1~GV in some hourly reconstructions. Such distortions may arise under different storm-time cutoff conditions. For example, \citet{2005JGRA..110.9S20B} noted that $R_{\rm c}$ can increase at storm onset, whereas \citet{2023JASTP.25206146G} discussed cases consistent with a reduction in cutoff rigidity. At the same time, the AMS daily proton data do not show a distinct FD-only trend in the fitted spectral parameters: FD days and quiet days follow the same overall parameter--flux relations in our appendix analysis. This suggests that the main FD-time anomaly in the hourly reconstruction is a localized cutoff-related distortion superposed on a broader day-scale spectrum that remains close to the usual AMS trend. Because these effects bias part of the spectrum during the disturbed interval, the published NM-based hourly proton series requires corrections in the affected bins and times for individual events.

The corrected product remains a constrained hourly reconstruction rather than an independent measurement. AMS daily fluxes are not inserted directly into the hourly series. Instead, AMS enters only through the daily spectral parameters and the selection of bins, while the hour-to-hour normalization is determined from the NM-based reconstruction in the remaining bins. No public proton data set currently provides both hourly resolution and comparable rigidity coverage for a direct cross-check, so the main evaluation criterion remains agreement of daily averages in the selected bins.

The physical origin of the localized deviations remains uncertain. A geomagnetic contribution to ground-based NM count rates during disturbed intervals is one plausible explanation, but event-specific spectral evolution and limitations of the learned mapping from NM observations to proton flux may also contribute. Separating these effects will require additional modeling and data. Within these limits, the corrected NM-based hourly proton flux provides a more consistent basis for studies of cosmic ray modulation on short timescales during disturbed intervals.

\clearpage
\appendix
\onecolumngrid
\setcounter{figure}{0}
\renewcommand{\thefigure}{A\arabic{figure}}
\renewcommand{\theHfigure}{A\arabic{figure}}
\renewcommand{\theHequation}{A\arabic{equation}}
\section{Technical Details}
\label{app:ams_constraint_details}

We use the daily AMS proton flux to describe how the proton spectrum changes with rigidity during ICME events. AMS does not provide an independent hourly spectrum in this analysis. Instead, the daily AMS spectrum supplies the spectral constraint used in the correction. This choice is physically motivated: AMS measures proton rigidity directly in space with event-by-event reconstruction and controlled exposure, whereas NM responses can be modified by storm-time changes in geomagnetic cutoff at fixed ground stations.

For two adjacent rigidity intervals with daily average proton fluxes $\Phi_1$ and $\Phi_2$, and the corresponding rigidity centers $R_1$ and $R_2$, we define the local spectral parameter as
\begin{equation}
\gamma =
\frac{\ln \Phi_2 - \ln \Phi_1}
{\ln R_2 - \ln R_1}.
\label{eq:gamma_local}
\end{equation}
The uncertainty of $\gamma$ is estimated from the flux uncertainties in the two adjacent rigidity intervals. For each daily AMS flux measurement, we combine the statistical error and the temporal error in quadrature. The uncertainty of the local spectral parameter is then written as
\begin{equation}
\sigma_{\gamma}
=
\frac{1}{\left|\ln R_2 - \ln R_1\right|}
\sqrt{
\left(\frac{\sigma_{\Phi_1}}{\Phi_1}\right)^2
+
\left(\frac{\sigma_{\Phi_2}}{\Phi_2}\right)^2
},
\label{eq:gamma_uncertainty}
\end{equation}
where $\sigma_{\Phi_1}$ and $\sigma_{\Phi_2}$ are the corresponding AMS flux uncertainties. Under this definition, the local spectral parameter represents the spectral slope between adjacent rigidity intervals for a given day.

These uncertainties in $\gamma$ are then carried into the fit of Eq.~(\ref{eq:gamma_fit}). The uncertainty of the corrected flux is calculated from the fit uncertainty of $\gamma(R)$ and propagated through Eq.~(\ref{eq:phirecon}).

The spectral-shape parameterization used in the main text is given by Eq.~(\ref{eq:gamma_fit}). Here we summarize the diagnostics that support its use during selected ICME intervals.

\begin{figure}[ht!]
\centering
\includegraphics[width=0.6\linewidth]{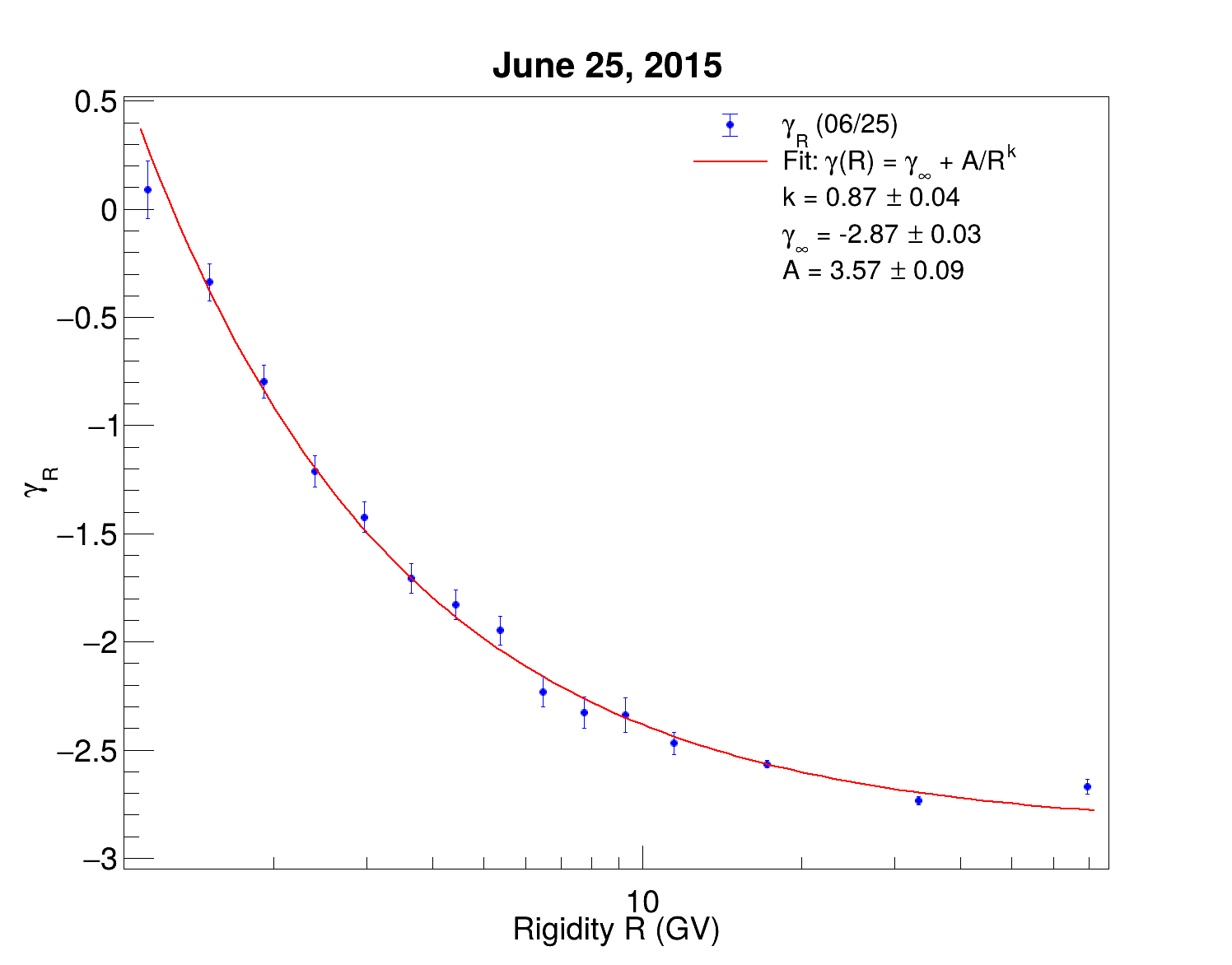}
\caption{Local spectral parameter $\gamma$ calculated from the AMS daily proton flux for the ICME event of 2015 June 25. The curve shows the fitted daily spectral parameter.}
\label{fig:gamma_fit}
\end{figure}

\begin{figure}[ht!]
\centering
\includegraphics[width=0.6\linewidth]{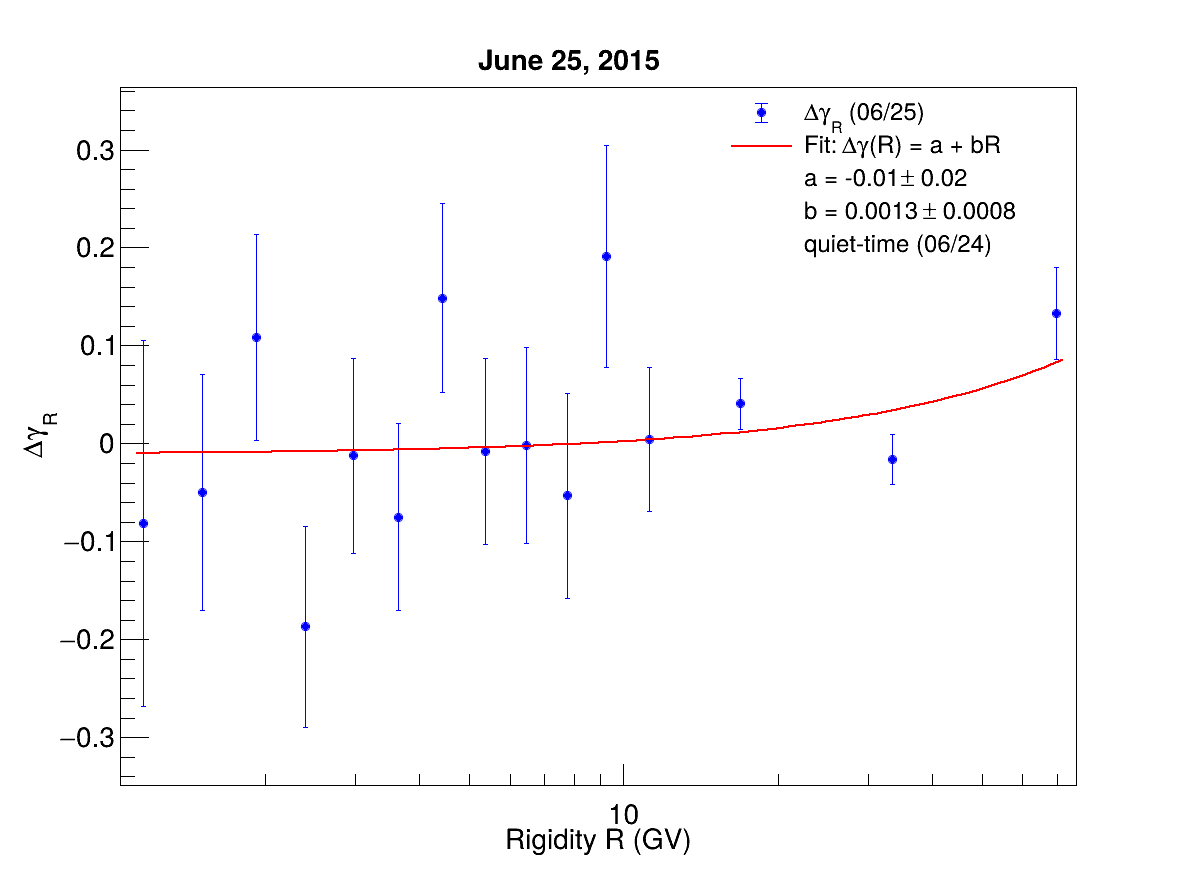}
\caption{Change in the local spectral parameter $\Delta\gamma$ calculated from the AMS daily proton flux for the ICME event of 2015 June 25. The reference is the spectrum on the quiet day before the event.}
\label{fig:delta_gamma}
\end{figure}

Figure~\ref{fig:gamma_fit} shows $\gamma(R)$ for the ICME event of 2015 June 25. Over the rigidity range relevant to the correction, the fitted curve follows the daily AMS spectral parameter. To examine how the spectral parameter changes during the event, we also define
\begin{equation}
\Delta\gamma(R,t)=\gamma(R,t)-\gamma(R,t_{\rm quiet}),
\label{eq:delta_gamma}
\end{equation}
where $t_{\rm quiet}$ denotes the quiet day before the event. Figure~\ref{fig:delta_gamma} shows the corresponding $\Delta\gamma$ relative to that quiet day.

\begin{figure}[ht!]
\centering
\includegraphics[width=\linewidth]{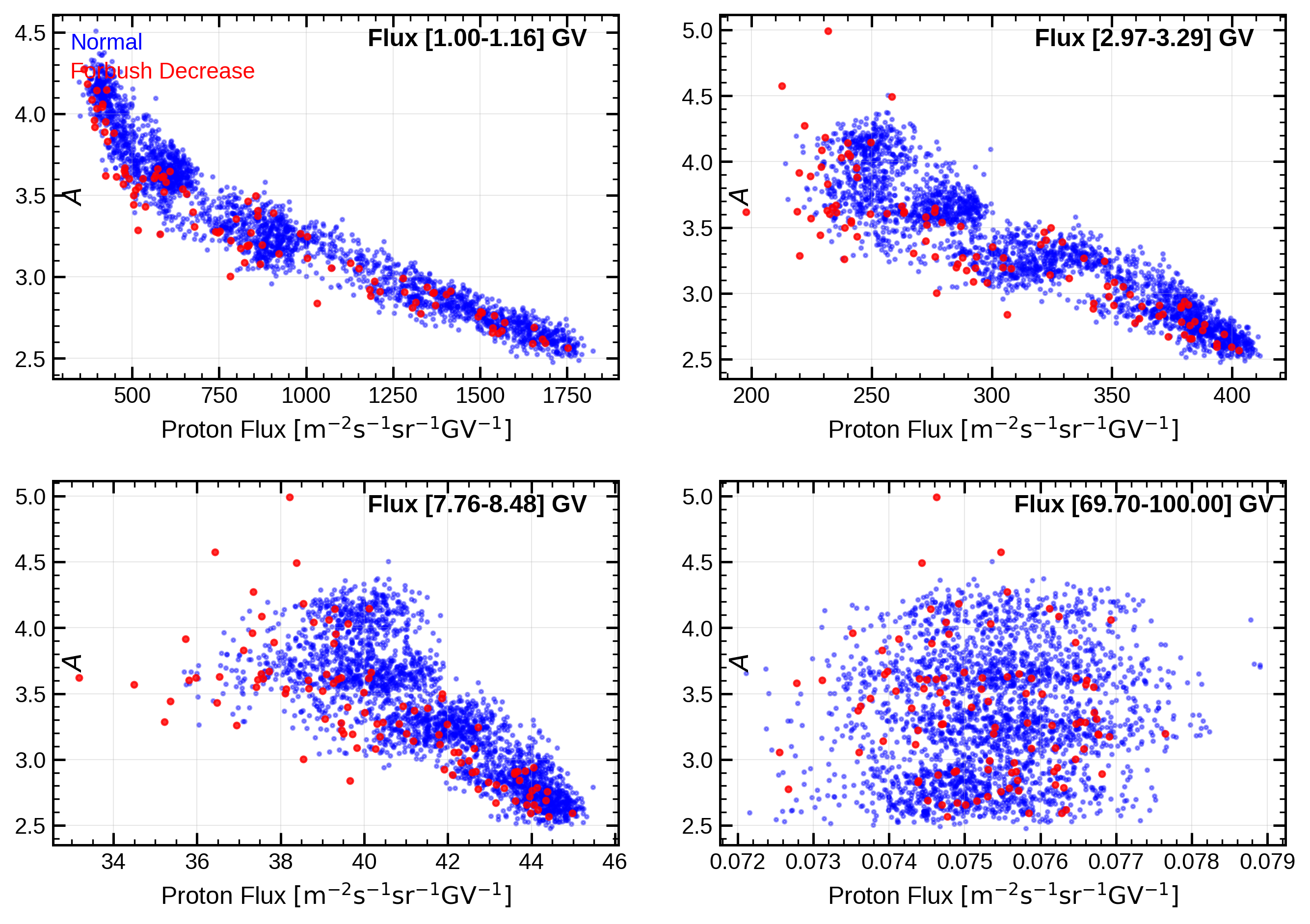}
\caption{Relation between the fitted parameter $A$ and the daily proton flux. Blue points denote quiet days, and red points denote days in FD events. The two groups show no clear separation.}
\label{fig:correlation_A_vs_flux}
\end{figure}

\clearpage
\begin{figure}[p]
\centering
\includegraphics[width=\linewidth]{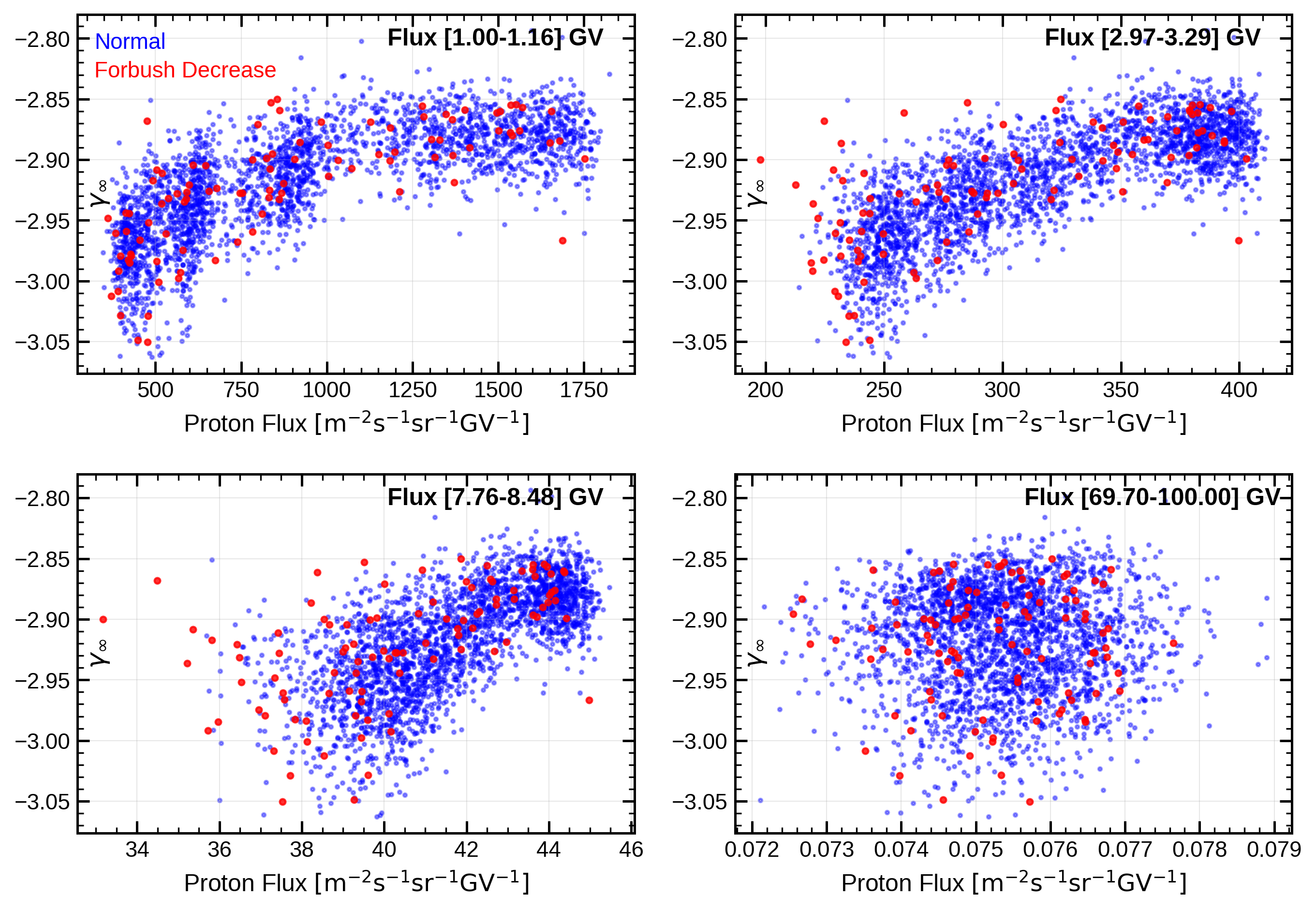}
\caption{Relation between the fitted parameter $\gamma_{\infty}$ and the daily proton flux. Blue points denote quiet days, and red points denote days in FD events.}
\label{fig:correlation_gamma_inf_vs_flux}
\end{figure}
\clearpage

\clearpage
\begin{figure}[p]
\centering
\includegraphics[width=\linewidth]{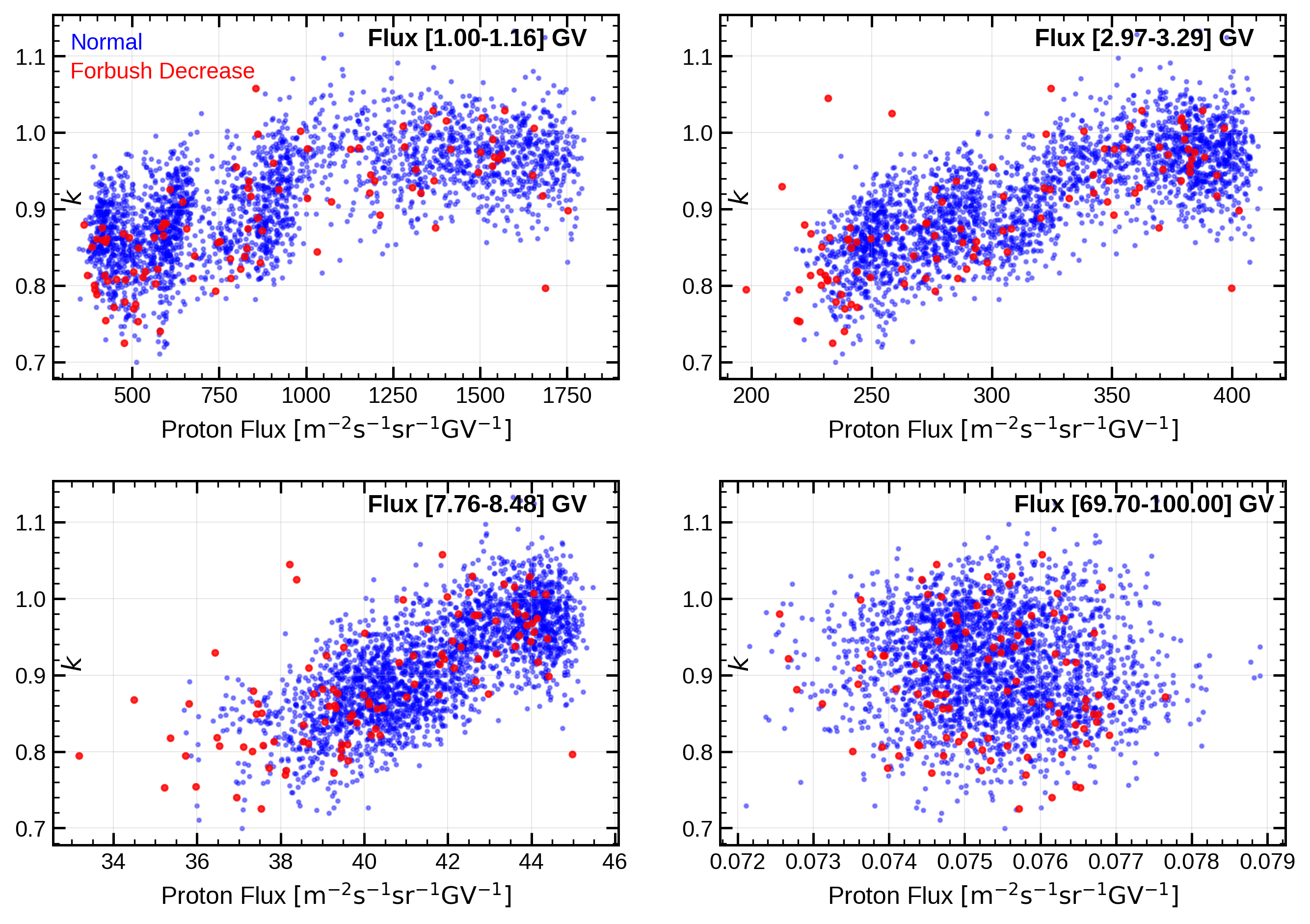}
\caption{Relation between the fitted parameter $k$ and the daily proton flux. Blue points denote quiet days, and red points denote days in FD events.}
\label{fig:correlation_k_vs_flux}
\end{figure}
\clearpage

In this event, the daily proton flux changes substantially through the ICME interval, but the fitted rigidity dependence keeps the same overall form. The variation of $\Delta\gamma$ remains small compared with the localized deviations seen in the hourly reconstruction.

We then compare the three spectral parameters in Eq.~(\ref{eq:gamma_fit}) with the daily proton flux for quiet days and for days in FD events. No clear separation appears between the two groups. Points from FD events shift toward lower flux values because the flux decreases during the FD phase, but the relations of $A$, $\gamma_{\infty}$, and $k$ with the daily flux remain similar. In other words, within the AMS daily proton data, FD days and quiet days follow the same overall parameter--flux trends rather than distinct spectral branches. This differs from the stronger spectral changes often inferred from NM-based studies (\citealt{2013SoPh..286..561A,2023JASTP.25206146G}). The corresponding correlations are shown in Figure~\ref{fig:correlation_A_vs_flux} for $A$, Figure~\ref{fig:correlation_gamma_inf_vs_flux} for $\gamma_{\infty}$, and Figure~\ref{fig:correlation_k_vs_flux} for $k$.

\begin{figure}[p]
\centering
\includegraphics[width=0.9\linewidth]{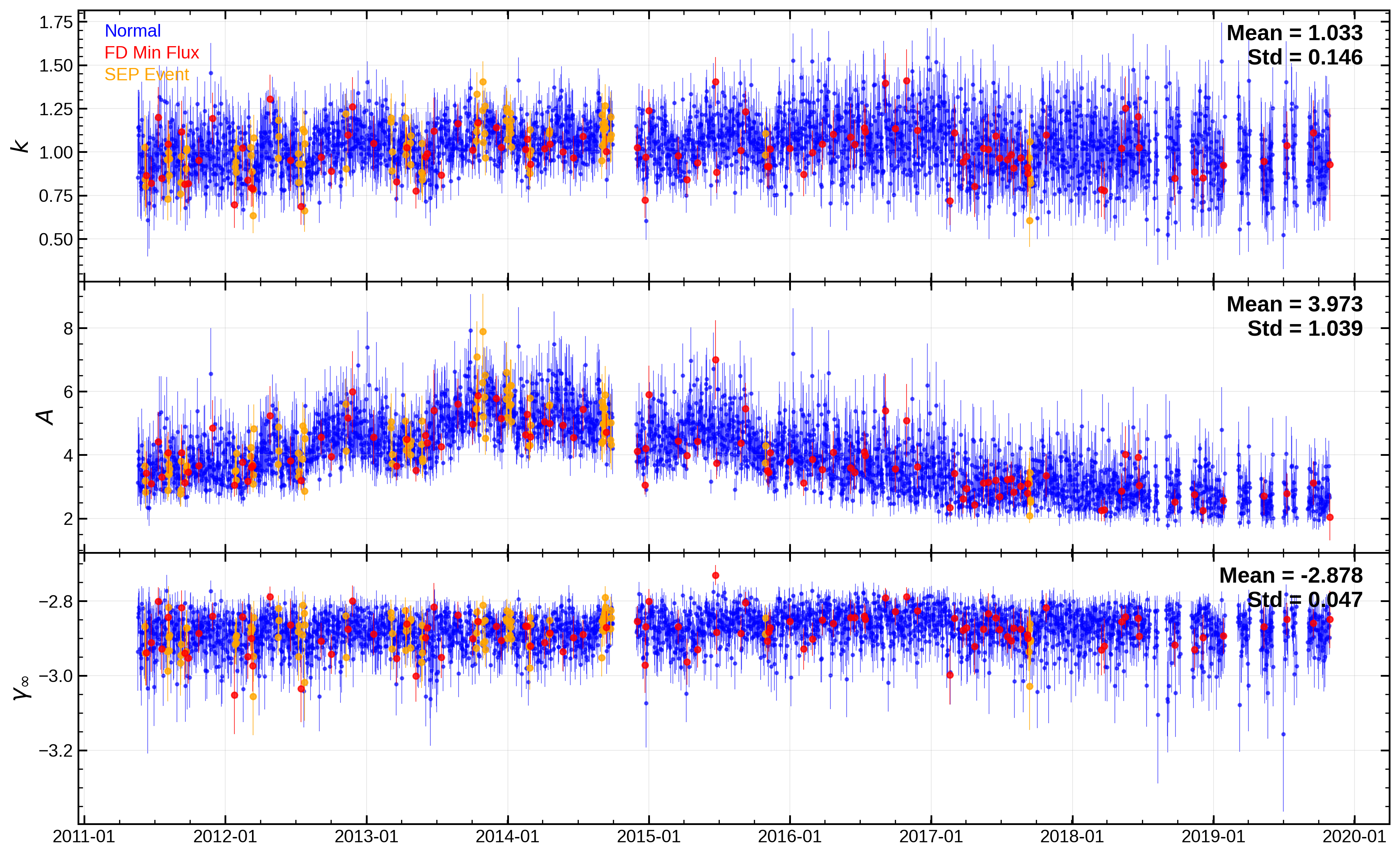}
\caption{Time evolution of the fitted spectral parameters $k$, $A$, and $\gamma_{\infty}$ for quiet days, days in FD events, and days affected by solar energetic particle events (SEPs). To compare the three categories, the fits are performed over a common rigidity range above 3~GV.}
\label{fig:params_longterm}
\end{figure}

We next compare the fitted parameters over the full daily sample. Figure~\ref{fig:params_longterm} shows the time evolution of $k$, $A$, and $\gamma_{\infty}$ for quiet days, days in FD events, and days affected by solar energetic particle events (SEPs). Because the SEP days exclude rigidities below 3~GV, the fits for all three categories are performed over a common rigidity range. The three groups show similar long-term trends.

\clearpage
\clearpage

Fits to all daily AMS spectra in the analyzed period from May 2011 through October 2019 give mean values and standard deviations of $-2.88 \pm 0.05$ for $\gamma_{\infty}$, $4.0 \pm 1.0$ for $A$, and $1.0 \pm 0.1$ for $k$. The scatter of these parameters remains limited over the full period. We therefore use the AMS daily spectrum as the spectral constraint during ICME events. In the correction step, the daily parameters are interpolated between neighboring days to provide a continuous hourly constraint and a smooth transition to the uncorrected series before and after the corrected interval.

The correction is restricted in both rigidity and time. For each ICME event, we define a detection window from the FD start day to the day of minimum flux; for one event, we also include the day before FD onset. Within this window, we compare daily averages in each rigidity bin with AMS and select bins for which the absolute daily-average deviation from AMS exceeds the AMS uncertainty at the $3\sigma$ level. For each continuous sequence of candidate days, the selected interval is extended by one day on both sides. Outside the selected bins and dates, the uncorrected hourly reconstruction is retained. Because no independent hourly proton measurements currently provide comparable rigidity coverage, both bin selection and assessment of the correction rely on agreement between daily averages and AMS. In what follows, we refer to the corrected series as the hourly reconstruction constrained by the daily AMS spectral fit.

\end{document}